\documentclass{emulateapj}
\usepackage{natbib}
\usepackage{rotating}
\usepackage{mathtools}
\usepackage{tablefootnote}

\begin{document}

\title{How many nucleosynthesis processes exist at low metallicity?}

\author{ C. J. Hansen}
\affil{Landessternwarte, ZAH, Heidelberg University, K\"onigstuhl 12, 69117  Heidelberg, Germany \\
\and  Dark Cosmology Centre, The Niels Bohr Institute, Copenhagen, Denmark}
\email{cjhansen@lsw.uni-heidelberg.de, cjhansen@dark-cosmology.dk}

\author{F. Montes}
\affil{Joint Institute for Nuclear Astrophysics, Michigan State University, East Lansing, MI 48824, USA\\
  \and National Superconducting Cyclotron Laboratory, Michigan State University,
   East Lansing, MI 48824, USA}
\email{montes@nscl.msu.edu}

\author{A. Arcones}
\affil{Institut f\"ur Kernphysik, Technische Universit\"at Darmstadt, Schlossgartenstr. 2,
Darmstadt 64289, Germany\\
\and GSI Helmholtzzentrum f\"ur Schwerionenforschung GmbH, Planckstr. 1, Darmstadt 64291, Germany}
\email{almudena.arcones@physik.tu-darmstadt.de}

\date{today}

\begin{abstract}
 
 Abundances of low-metallicity stars offer a unique opportunity to understand the
contribution and conditions of the different processes that synthesize heavy
elements. Many old, metal-poor stars show a robust abundance pattern for elements
heavier than Ba, and a less robust pattern between Sr and Ag.  Here we probe if two
nucleosynthesis processes are sufficient to explain the stellar abundances at low metallicity, and we
carry out a site independent approach to separate the contribution from these two
processes or components to the total observationally derived abundances.  Our approach
provides a method to determine the contribution of each process to the production of
elements such as Sr, Zr, Ba, and Eu. We explore the observed
star-to-star abundance scatter as a function of metallicity that each process leads to. Moreover, we use the deduced
abundance pattern of one of the nucleosynthesis components to constrain the
astrophysical conditions of neutrino-driven winds from core-collapse supernovae.

\end{abstract}

\keywords{Galaxy: evolution – Galaxy: stellar content – nuclear reactions, nucleosynthesis, abundances – stars: abundances – supernovae: general}
%----------------------------------------------------------------------------------------------------
\section{Introduction}
\label{sec:intro}
%----------------------------------------------------------------------------------------------------
Observations of stellar abundances at low metallicities are necessary
to understand the origin of elements heavier than iron \citep{Spite1978,Truran1981,Ryan1996,burris00,fulbright02,Honda2004,Barklem2005,francois,Lai08,hansen12,Yong2013, roederer2014}. At early times, those elements originated from one or more primary processes, which
imply that no seed nuclei needs to be produced prior to the
nucleosynthesis process because the process can create the seeds
itself. An example of this is the r process; seed nuclei are synthesized
starting with neutrons and protons by charged-particle reactions combined with neutron captures. Note that the s process is
also observed at low metallicities even if it is a secondary process,
however, such observations are mostly related to binary systems (see, e.g.,
\citealt{beers2005,masseron2010,bisterzo11} for a discussion).

\begin{figure}
  \begin{center}
    \includegraphics[width=0.98\linewidth]{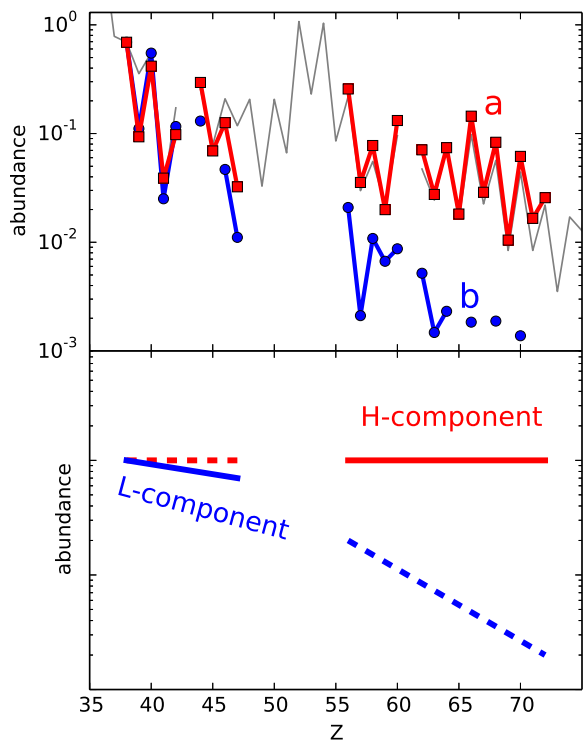}
  \end{center}
  \caption{Two metal-poor stellar abundance patterns are shown in the
    upper panel compared to the solar r process (grey thin line). The
    pattern ``a'' (red line and squares) corresponds to CS~22892-052,
    and the ``b'' pattern (blue line and circles) to HD~122563. All
    abundances are normalized to Sr. The bottom panel shows a
    sketch of the components that may contribute to produce these two
    patterns. The upper one is the H component (red line) and the
    lower one the L component (blue). The solid lines indicate that elements in this range are mainly produced by this component, while the
    dashed lines indicate that the component may not pay the predominant contribution to this elemental range.}
  \label{fig:LH-components}
\end{figure}

The r process seems to be robust based on observations of
old r-II stars\footnote{see \citet{beers2005} for a definition}, which show an enrichment in heavy elements ($Z>50$),
see \cite{Cowan.etal:1995,Hill2002,cowan02,Sneden2003}, and
\cite{Sneden.etal:2008} for a recent review. However, there are
several indications that this robustness cannot be generalized to
all elements. Independent indications of several different
processes were initially found in meteorites
\citep{Wasserburg.etal:1996}. Here, we highlight two additional indications. First, the
abundance pattern between Sr and Ag varies among stars, even
if those have a very uniform pattern for heavy elements
\citep[e.g.,][]{Sneden.etal:2008, hansen12, hansen14}. Second,
improved observations have demonstrated that the abundances for the
heaviest elements ($Z>50$) can vary significantly compared to the
$38<Z<50$ elements, as shown in Fig.~\ref{fig:LH-components} (see also
\citet{Aoki2005,Roederer2010}). This lack of robustness has raised the
question: how many primary processes contribute to the abundances
observed at low metallicities?

Here, we provide a possibility to explain the variety in stellar
abundances at low metallicities and trace individual processes, an option that only isotopic
abundances otherwise offer. Our study focuses on the abundance patterns from a large sample of metal-poor stars. We follow the nomenclature of \cite{Qian.Wasserburg:2001,
  Qian.Wasserburg:2007} using an H- and L-component to explain the
formation of the heavy ($Z>50$) and lighter heavy ($38<Z<50$)
elements, respectively.  
We adopt a simple approach to explain the observationally derived abundances in metal-poor stars; only two nucleosynthesis contributions (the H- and L-components) are needed to reproduce the stellar abundances within an estimated uncertainty ($\pm0.32$\,dex). In practice, we use a linear superposition of the H- and L-component (see also \cite{Li2013}) to explain the stellar abundance pattern from the sample compiled by \citet{frebel10} (after applying five selection criteria to remove contamination from s processes, self-pollution due to stellar mixing processes, etc.). These assumptions, though simple, are sufficient to explain the stellar abundance patterns from observations for most of the metal-poor stars passing the selection criteria.

There are several nucleosynthesis processes/astrophysical sites that are H- and L-component candidates.
The H component is most likely the r process that produces
heavy elements up to U via rapid neutron captures compared to beta
decays. Although this process was already proposed in 1957
\citep{Burbidge.Burbidge.ea:1957}, there are still open questions
concerning the astrophysical site and the neutron-rich nuclei involved
\citep{arnould.goriely.takahashi:2007}. The best studied site (after
the work of \cite{Woosley.etal:1994}) is core-collapse supernovae and
their neutrino-driven winds. However, the conditions reached in these
environments are not sufficiently neutron rich to produce heavy elements
up to uranium \citep[see][and references
  therein]{arcones.thielemann:2013}. Neutrino-driven winds may be
slightly neutron rich or even proton rich \citep{ Roberts.Reddy:2012,
  MartinezPinedo.etal:2012}. Another possibility to produce the
heaviest elements in core-collapse supernovae are explosions driven by
magnetic fields \cite[see, e.g.,][]{Winteler.etal:2012}. Mergers
of two neutron stars or a neutron star and a black hole are also promising
candidates to produce heavy r-process elements
\citep{Lattimer.Schramm:1974, Freiburghaus.Rosswog.Thielemann:1999,
  Korobkin.etal:2012, Bauswein.etal:2013, Hotokezaka.etal:2013}.

While the H component is associated with the r process, the
L component may be one of several processes. One
possible site to produce the L component is neutrino-driven wind in core collapse supernovae. This possibility is explored in this paper. In
these events, the alpha process \citep{Woosley.Hoffman:1992,
  Witti.etal:1994, Hoffman.Woosley.etal:1996} or charged-particle
reactions \citep[CPR;][]{Qian.Wasserburg:2007} produce seed
nuclei. Later, if the conditions are slightly neutron rich, a weak
r process can form elements up to $Z\sim50$ (see \cite{Fara10,
  arcones.bliss:2014}). In proton-rich conditions, the $\nu p$ process can also
reach those nuclei \citep{Froehlich06, Pruet.Hoffman.ea:2006,
  Wanajo:2006}. In both cases, it is possible to produce an abundance
pattern similar to the L component \citep{Arcones.Montes:2011,
  Wanajo.Janka.Mueller:2011}. Therefore, the L component may be the
weak r process or $\nu p$ process, or even a combination of both. Note
that these processes may also occur in neutron star mergers
\cite{Perego.etal:2014, Just.etal:2014, Metzger.Fernandez:2014}.

The L component could also be what \cite{Travaglio.etal:2004} called
LEPP for Lighter Element Primary Process, and it was used to explain the contribution
of a process to the abundances from Sr to Ag (see also
\cite{Montes.etal:2007} and \cite{hansen12}). Although the initial motivation of the LEPP was to explain missing solar abundances using stellar models, it has been pointed out by, e.g., \citet{trip2014} that such a process may not be needed after all. Nevertheless, additional possibilities to produce an
  L component may be a primary s process in fast rotating stars
  \citep{Frischknecht.etal:2012, Pignatari.etal:2008}, or an
  early s process leading to a mass
  transfer in an extremely metal-poor binary star system \citep{straniero04,lucatello05,masseron2010,bisterzo10,bisterzo11,stancliffe11,Cruz2013}.

In this paper, we extract the L and H components from the metal-poor stellar abundance patterns
as indicated in Fig.~\ref{fig:LH-components}. As such, the extracted abundance pattern is site-independent. Although we explore neutrino-driven winds as an L component possibility, we do not exclude other processes/sites from being suitable candidates. The
pattern ``a'' (red line and squares) corresponds to \object{CS~22892-052} \citep{Sneden2003} with an
enhancement of heavy elements and a robust pattern compared to the
solar scaled pattern. The
``b'' pattern (blue line and circles) corresponds to \object{HD~122563} \citep{george1963,Honda2004,Honda2007} and is characterized by higher
abundances of the lighter heavy elements ($Z<50$) and much lower
abundances for $Z>50$. 

 This paper is organized as
follows. The large sample of metal-poor stars from \citep{frebel10} is reduced, homogenized,
and discussed in detail in Sect.~\ref{sec:obs_set}. In
Sect.~\ref{sec:components} the methods to obtain the H  and
L components are introduced and compared to observations. In
Sect.~\ref{sec:results} the separated L abundances are used to place
constraints on a possible formation site; the neutrino-driven winds. Summary and conclusions can
be found in Sect.~\ref{sec:summ}.

%----------------------------------------------------------------------------------------------------
\section{Data---Observationally derived stellar abundances}
\label{sec:obs_set}
%----------------------------------------------------------------------------------------------------
To minimize the number of contributing processes, we   only consider metal-poor (old) stars that are less enriched by s processes compared to younger (more metal-rich) stars. To clean the sample presented in \citet{frebel10}, we apply five selection criteria (below) to the large, inhomogeneous sample:

\begin{enumerate}
\item $[$Fe/H$] < -2.5$: this cut removes the majority of the s-process
  contribution. \citet{Travaglio.etal:2004} state that the s process(es)
contributes little to halo stars below [Fe/H]=$-1.5$. Recently,
\citet{hansen14} found that the s process yields may be traceable below
      [Fe/H]=$-1.5$ possibly down to $-2.5$, but not lower.  Hence,
      neglecting the s process below [Fe/H]$=-2.5$ is an observationally justified
      assumption.
 Including stars with a metallicity up to $-2.5$ (instead of staying below $-3$)
  ensures that a large fraction of heavy elements are still
  detectable, even if the star is not extremely enhanced in neutron-capture elements.
  Most absorption lines weaken as metallicity
    decreases, making the abundance analysis increasingly difficult with decreasing metallicity. 
\item $[$C/Fe$] <0.7$: this ensures that no carbon enhanced metal-poor
(CEMP) star is included \citep{aoki2007}, as most   Fe-poor stars are carbon-enhanced stars, many of which are also s-process enriched. 
\item $[$Ba/Fe$] < 1.0$: this cuts out Ba stars and together with the
  previous criteria removes strong s-process enhancement in, e.g., CEMP-s stars.
\item Excluding abundances that are only upper limits yields a better and more solid final
  abundance pattern with known reasonable sized uncertainties. This
  facilitates a more direct comparison of observations   and predictions. 
\item   Most heavy elements are detected in the extended atmospheres of giant stars, that due to stellar evolution have had their surface composition altered. This change is normally seen in their carbon and nitrogen abundances, which is why we place the cuts: $[$C/N$] <-0.4$ and [N/Fe] $>0.5$. These exclude stars with
  internal mixing owing to the stellar evolution \citep{spite05}. Very evolved stars (giants) burn C into   N and later O, which will result in lower C and higher N and O abundances.
\end{enumerate}

The original sample from \citet{frebel10} is a compilation of
different sources from the literature. Therefore, our reduced sample
is inhomogeneous owing to the variety of different stellar parameter
scales and methods used to derive these abundances.
After carefully examining the observational data, inconsistencies
between the   ``original'' data found in the literature and the compilation
in \citet{frebel10} were revealed for two of our reduced sample's
stars. As a consequence, these were removed from our final sample.  The
star \object{CS~30325-094} has a Eu abundance that is observed only as an upper
limit in \citet{francois} and the Pm abundance was not found in the
quoted reference. In addition, \object{CS~22783-055}
\citep{McWilliam1995,McWilliam1995b} has abundances in the \citealt{frebel10} table that
we were not able to find in the literature.

If we also require that each star needs to have at least
five heavy element detections or more (i.e., we do not count upper
limits), the final reduced sample consists of 39 stars. However, if we
loosen criteria (5) to only affect $[$C/N$] <-0.4$, and set no [N/Fe]
constraint, the sample is increased to 53 stars (model 4 in
Table~\ref{tab:FitTable}).

The   final abundance pattern\footnote{The $\log \epsilon$ abundances were adopted and
when necessary converted to relative abundances using the solar
abundances from \citet{AndersGrev}. }
 of each star, consists of both neutral and
ionized elemental abundances, and some species (e.g., Sr I---the
minority of Sr) are more affected by   nonlocal thermodynamic equilibrium (NLTE and possibly three-dimensional (3D))
effects than other species (e.g., Sr II---the majority of
Sr). This introduces a possible bias between the mixture of
neutral/ionized elements that compose the total abundance pattern, and
this bias may exceed the uncertainties stemming from the inhomogeneity
of the sample. The NLTE corrections are not calculated for all
the heavy elements (owing to the lack of atomic physics), and even fewer of the
heavy elements have NLTE corrections calculated for a large stellar parameter
space (some of these elements are Sr and Ba which have been
investigated in detail,
e.g., \citealt{Andrievsky2009,Andrievsky2011,Bergemann2012,hansen13}). Fortunately,
several of the heavy elements have abundances derived from the majority
species (which, in many cases, are single ionized lines), and for some
of these elements the 3D and NLTE effects may cancel out, thereby
removing the bias in the heavy element abundance pattern.

We account for the sample inhomogeneity that may lead to biases in the abundance pattern by
propagating slightly increased uncertainties into the nucleosynthesis
components in the next sections
(Sect.~\ref{sec:components} and Appendix \ref{sec:appendix}).
%----------------------------------------------------------------------------------------------------
\section{Nucleosynthesis components}
\label{sec:components}
  Following \cite{Qian.Wasserburg:2001, Qian.Wasserburg:2007,
  Qian.Wasserburg:2008}, we call the two components that contribute to the metal-poor stars the L component
and H component (see Fig.~\ref{fig:LH-components} for schematic
representation). We assume that the H component is the main source of
heavy r process elements ($Z>50$), but it may also contribute to the
lighter heavy elements ($38<Z<50$). In contrast, the L component
contributes mainly to these lighter heavy elements, but may also extend toward the heavy $Z>50$ elements. With these two components, there are several
possibilities to explain the typical patterns shown in the upper panel
of Fig.~\ref{fig:LH-components}. The pattern ``a'' can be produced only by an H component going from Sr to the heaviest elements. Another possibility to explain the ``a'' pattern would be a
combination of an H component that contributes only to the heavy part
($Z>50$) and a L component to explain the pattern below $Z<50$. The ``b'' pattern can be the result of a single
L component that extends toward heavy elements or the combination of
an L component up to $Z=50$ and a small contribution from the H component to
explain the low abundances of heavy elements. 

\subsection{Component identification}
\label{sec:comp_identification}
%----------------------------------------------------------------------------------------------------

In order to explain the abundances of our reduced sample of stars
described in Sect.~\ref{sec:obs_set} by an L  and H component, we first
extract the abundance pattern of these two components.  The pure
nucleosynthesis component patterns are obtained using three different
methods (M1, M2, M3) in order to test their robustness and to estimate
their uncertainties (see Appendix~\ref{sec:appendix} for
details). All methods use abundances from  the metal-poor
stars HD~122563 and HD~88609 (which have large [Sr/Eu] ratios,
\citealt{Honda2007}) and  CS~22892-052 (which has a
large [Eu/Fe] ratio, \citealt{Sneden2003}).

\begin{figure}[!htb]
  \begin{center}
    \includegraphics[width=0.99\linewidth]{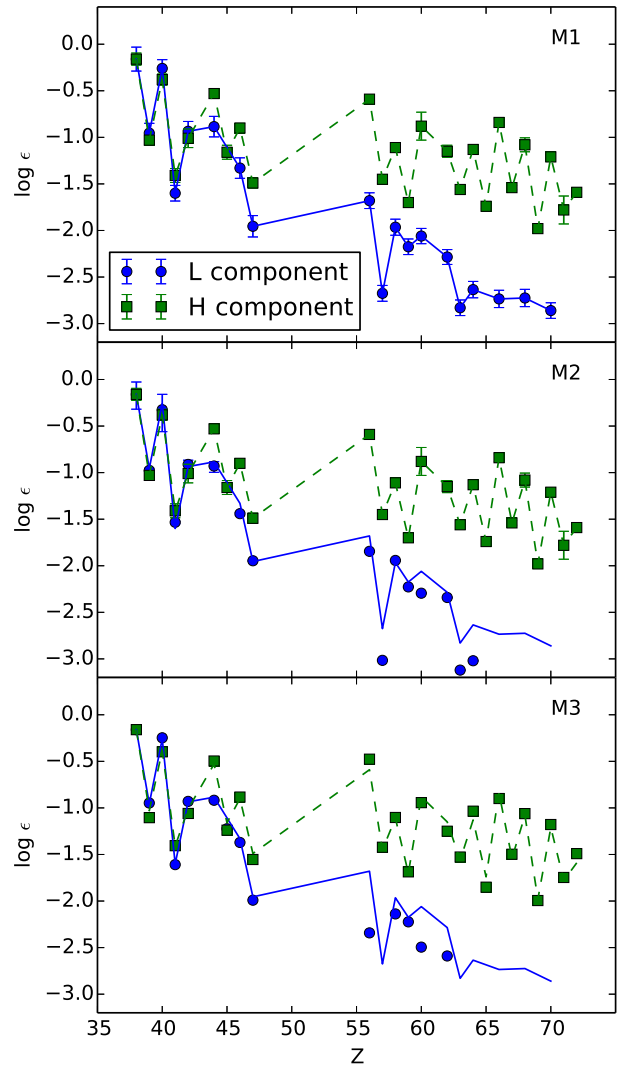}
  \end{center}
  \caption{Pure L component and H component obtained using the three
    methods described in the text. In all three panels, the solid blue (dashed green)
    line corresponds to the L(H) component obtained in method~1 (M1), while the
    symbols vary with the methods.}
  \label{Fig:PureComponents}
\end{figure}

Method one (M1) assumes that HD~122563\footnote{the same applies for
  \object{HD~88609}} has only been enriched by the L component (due to
the large Sr-enrichment) while \object{CS~22892-052} has only been
enriched by the H component (due to its large Eu-enrichment). As such,
their abundances already show the individual components. These are
shown in the upper panel of Fig.~\ref{Fig:PureComponents}.

 Method two (M2) follows \cite{Montes.etal:2007} and assumes that
 while \object{CS~22892-052} has a pure H-component abundance,
 \object{HD~122563} shows a large L component combined with a small
 H-component contribution.  This small contribution is removed by
 subtracting the abundances of \object{CS~22892-052} from the
 \object{HD~122563} abundance (by scaling to the average of Eu, Gd,
 Dy, Er, and Yb abundances in each star). This leaves only the pure
 L-component pattern as shown in the middle panel of
 Fig.~\ref{Fig:PureComponents}.

Method three (M3) follows \cite{Li2013} and assumes that the mentioned
metal-poor stars' abundances do not have pure nucleosynthesis
components, but instead have a dominant contribution from one of the
components. The pure component abundances are obtained by
systematically eliminating the abundances from the process that
contributes the least.  The first-order L-component abundance is
obtained by subtracting the CS~22892-052 abundances (scaled to the Eu
abundance which is predominantly produced by the H component) from the
average of the HD~122563 and \object{HD~88609} abundances. Conversely,
the first-order H-component abundance is obtained by subtracting
the average of the HD~122563 and HD~88609 abundances (scaled to the Fe
abundance) from the CS~22892-052 abundances. The individual
components are obtained by further subtraction of the remaining
contribution, e.g., the $n^{th}$-order L-component abundance is
obtained by subtracting the $(n-1)^{th}$-order H-component abundances
scaled to Eu from the average of the \object{HD~122563} and \object{HD~88609} stellar
abundances. The procedure is repeated until the differences in both the
L component and H component are smaller than the observational
error. Figure~\ref{Fig:PureComponents} (bottom panel) shows the
nucleosynthesis components obtained following this method. The
robustness of the derived components   using method three was checked by using different
combinations of metal-poor stars.  Since the assumption is that all
stellar abundances at low metallicity have contributions from robust L 
and H components, any pair of metal-poor stars could, in
principle, be used to obtain the pure components (see
Appendix~\ref{sec:appendix} for additional tests). However, the errors
of the iterative method are smaller when the differences between the
observed abundances are larger.

\begin{figure}[!hb]
  \begin{center}
    \includegraphics[width=0.5\textwidth]{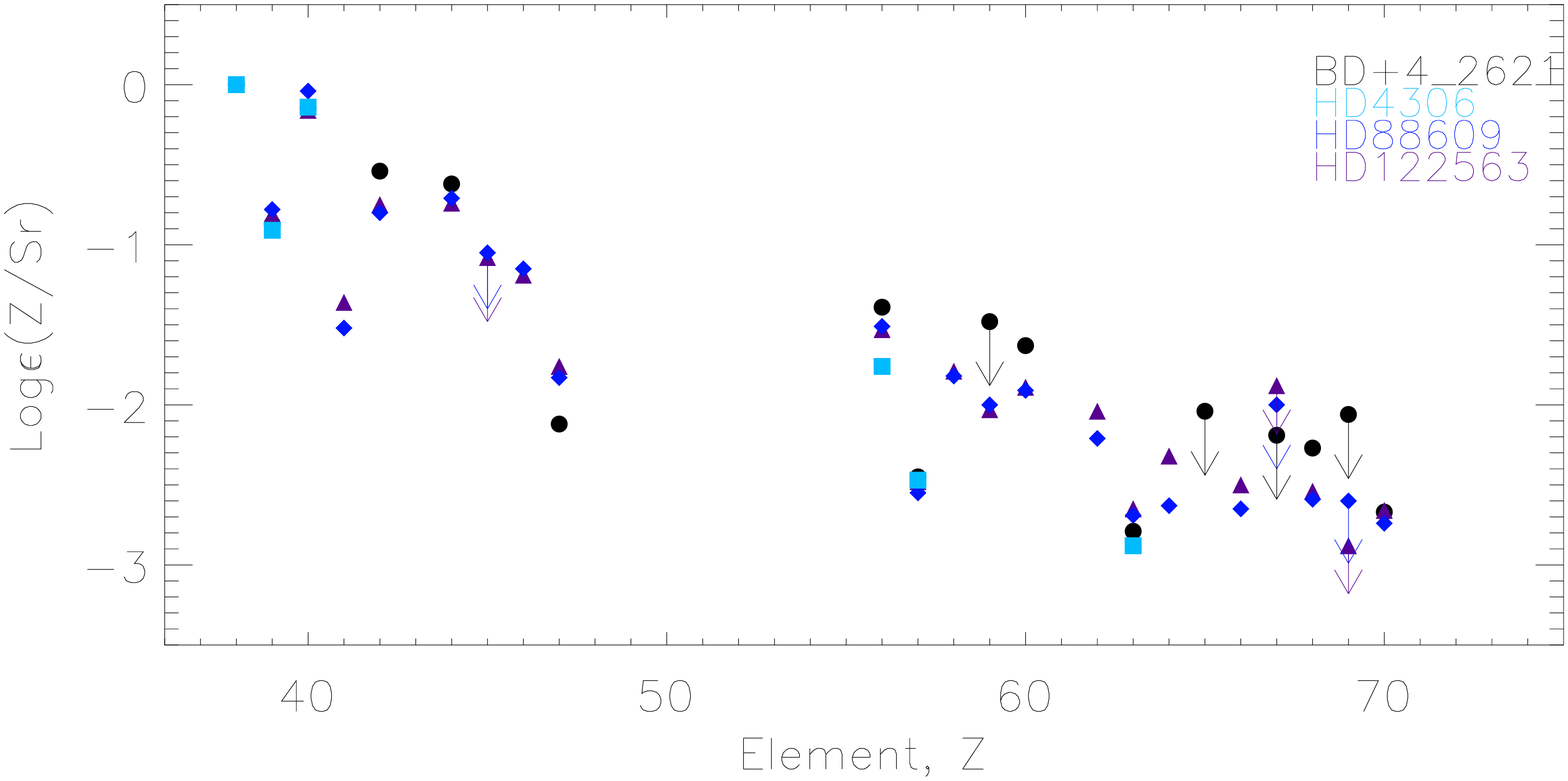}
  \end{center}
  \caption{Abundances normalized to Sr for four well-known L-component stars.}
  \label{fig:L-component}
\end{figure}

The derived H-component abundances shown in
Fig~\ref{Fig:PureComponents} are remarkably consistent between the
different methods. The calculated abundance difference between methods
is within $\pm 0.2$\,dex for every element. In contrast, the derived
L-component abundances vary up to an order of magnitude for elements
heavier than Ba. For elements between Sr and Ag, the obtained
L-component abundances are within $\pm 0.2$~dex for all methods.  In
the following, we assume that both components have an uncertainty of $\pm
0.2$\,dex for every element and consider the possibility, that the
L component may be limited to elements up to Ag (see
Sect.~\ref{sec:fitting}).  Furthermore, we assume that the components are
``robust'' within the abundance uncertainties. This is a fair
assumption for the H component, but may not be true for the
L component as shown in Fig.~\ref{fig:L-component}. The stars
presented in this figure have a typical L-component pattern that is
fairly consistent (within $\sim 0.2$\,dex) for Z$<$50. However, we
refer to Sect.~\ref{sec:scatter} and Fig.~\ref{fig:L-componentall} for
further discussion on the robustness of the L component.    We
  stress that the robustness of the L component is an assumption that
  we will test in the following sections.

%----------------------------------------------------------------------------------------------------
\subsection{Fitting observations with two components}  %3.2
\label{sec:fitting}
%----------------------------------------------------------------------------------------------------
The H and L components introduced in the previous section are  
  assumed to be responsible for the abundances ($Y$) of metal-poor stars. For every element with $Z>50$,
the abundance of the sample stars can be expressed as a
combination of these components ($Y_{H}$ and $Y_{L}$):
\begin{equation}
Y_{calc}(Z) = \big(C_{H} Y_{H}(Z)+C_{L} Y_{L}(Z) \big)\times10^{[\mathrm{Fe/H}]},
\label{eq:Ycalc}
\end{equation}
where $C_{H}$ and $C_{L}$ are the weights of the H and L components to
the abundances of the star, respectively. It should be noted that
since there is an arbitrary scaling factor when defining $Y_{H}(Z)$
and $Y_{L}(Z)$, the values of $C_{H}$ and $C_{L}$ are relative and
only their overall trends have physical significance. The factor
$10^{[\mathrm{Fe/H}]}$ is introduced to normalize the abundances at
different metallicities.

In order to find the coefficients $C_{H}$ and $C_{L}$ that best match
the observationally derived abundances in metal-poor stars, the
following $\chi^{2}$-distribution was minimized,
\begin{equation}
\chi^{2}= \frac{1}{\nu} \sum_{Z_{\mathrm{range}}} \big(\log Y_{\mathrm{observed}}(Z) - \log Y_{\mathrm{calc}}(Z)\big)^2/ \Delta(Z)^2,   
\label{eq:fittedchi}
\end{equation}
where Z$_{\mathrm{range}}$ is the elemental range considered in the
minimization, $\Delta(Z)$ corresponds to the abundance uncertainty of
element $Z$ from both the observation and the nucleosynthesis
component determination, and $\nu$ is the number of degrees of freedom
in the fit (number of elements observed in Z$_{\mathrm{range}}$ minus
the number of fitted coefficients $C$). The uncertainty in the
observation (0.25\,dex) and the intrinsic error in the component
estimation (0.2\,dex) were added in quadrature to obtain $\Delta$(Z) =
0.32\,dex for all elements, see Appendix~\ref{sec:appendix}. 

Once the $\chi^{2}$-distribution (Eq.~(\ref{eq:fittedchi})) has been
minimized for a given star, the minimum $\chi^{2}$ obtained with the
preferred $C_{H}$ and $C_{L}$ coefficients is associated with that
star. A star that has its abundances well (badly) calculated within
this approach, would result in a low (high) $\chi^{2}$ value. A large
number of stars with high $\chi^{2}$ values would indicate that the
assumptions in our approach are incorrect. 

\begin{figure}[!htb]
  \begin{center}
    \includegraphics[width=0.99\linewidth]{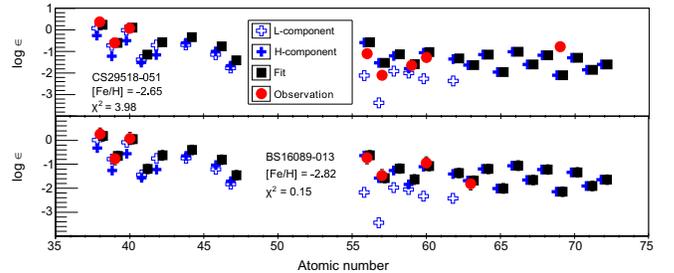}
  \end{center}
  \caption{Poor (top) and good (bottom) $\chi^2$ fit to two sample stars.}
  \label{Fig:x2multi}
\end{figure} 

We calculate the $\chi^{2}$-distribution
(Eq.~(\ref{eq:fittedchi})) for the stars in our sample. We have studied
various models that assume different elemental ranges for every
component and use M3 for the H component and vary the method for the
L component. These models are summarized in Table~\ref{tab:FitTable}.
In model 4, the number of stars in the sample is larger and marked
with an asterisk to indicate that stars in this model fulfill criteria
1 to 4, but not 5   (the criteria are described in Sect.~\ref{sec:obs_set}). 

Stars with low $\chi^{2}$ values show a typical trend (an example of
this is shown in the lower panel of Fig.~\ref{Fig:x2multi}): Z$\ge$56
abundances fitted by the H component, and 38$\le$Z$\le$47 abundances
fitted by a combination of L and H components. Thus, using the
L component resulting from different methods has no effect on the
fit because variations among methods are significant only for elements
with $Z>50$.  In addition, in the range 38$\le$Z$\le$47, the H- and
L-component abundances are remarkably similar and within the error bar
$\Delta$(Z) used in the fit. Since the L component is not making
  significant contributions to Z$>56$   (as can be seen from the good fits using models 1 and 2) a combination of L- and H-component
contributions in the range of 38$\le$Z$\le$47, is almost equivalent to
having a single process making only those abundances (Z$\le$47)
independently of the process responsible for the Z$\ge$56 abundances   (model 3 in Table \ref{tab:FitTable}).
 
\begin{table}[!htb]
\caption{Details (component, sample size, and $\chi^2$) of the four models.}
\begin{center}
\begin{tabular}{|r|c|c|c|c|}
\hline
           &  H-                  & L-                      &   Num. of                  & Stars with  \\
model  & component      &    component      &  stars in sample  & $\chi^{2} \ge \chi^{2}_{5\%}$\\
\hline
1  &    M3 		& M3 		 &  39 stars 	     & $<$1\% \\
2  &    M3 		& M1 		 &  39 stars 	     & $<$1\% \\
3  &    M3, Z$>$47 	& M3, Z$<$56    &  39 stars            & $<$1\% \\
4  &    M3 		& M3 	         &  53 stars$^{*}$   & 11\% \\
\hline
\end{tabular}
\label{tab:FitTable}
\end{center}
\end{table}

In order to test how well our models fit the stellar abundances, the
expected $\chi^{2}$-probability distribution was calculated by adding
up the expected $\chi^{2}$-distribution of every star considered. Each
star has an expected $\chi^{2}$-distribution that depends only on the
degrees of freedom. The expected $\chi^{2}$-probability distribution
can thus be expressed as
\begin{equation}
f(\chi^2) = \sum_{i} \frac{\nu_i}{2^{\nu_i/2} \Gamma(\nu_i/2)} e^{-\chi^2 \nu_i/2} (\chi^2 \nu_i)^{(\nu_i/2-1)},
\label{eq:expectedchifunction}
\end{equation}
where $\nu_i$ is the number of degrees of freedom for star $i$.  The
$\chi^{2}$ test relies on the assumption that each elemental abundance
is normally distributed within a given error. If the expected
$\chi^{2}$-probability distribution is too large compared to the
minimum $\chi^{2}$ values satisfying Eq.~(\ref{eq:fittedchi}), we either
conclude that a statistically improbable excursion of $\chi^2$ has
occurred, or that our model is incorrect.  If, on the other hand, the
expected $\chi^{2}$-probability distribution is too small, it is not
indicative of a poor model, but that a statistically improbable
excursion of $\chi^2$ has occurred, or that $\Delta$(Z) has been
overestimated in the model.

\begin{figure}[!htb]
  \begin{center}
    \includegraphics[width=0.99\linewidth]{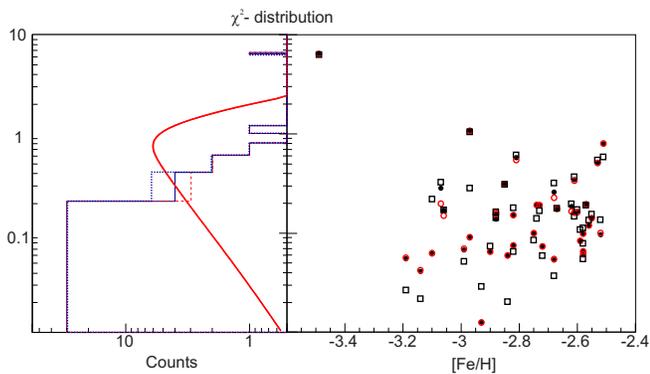}
  \end{center}
  \caption{Minimum $\chi^2$ values as a function of metallicity
    are shown in the right panel. Left panel shows expected
    $\chi^2$-distribution using Eq.~(\ref{eq:expectedchifunction}) (red line) and
    histograms of the $\chi^2$-distribution values for different models
    presented in Table~\ref{tab:FitTable}. Model~1 is shown as a solid
    line (left panel) and black solid circles (right panel), model~2
    as a dashed line (left panel) and red open circles (right panel),
    and model~3 as a dotted line (left panel) and black open squares
    (right panel).}
  \label{fig:mod6_0.32_5}
\end{figure}

The minimum $\chi^{2}$ is shown as a function of metallicity in the
right panel of Fig.~\ref{fig:mod6_0.32_5} for the first three models
listed in Table \ref{tab:FitTable}.  In the left panel, one can see
the expected $\chi^2$-distribution using Eq.~(\ref{eq:expectedchifunction}) (red
line) and histograms of the $\chi^2$-distribution values based on
models 1, 2, and 3 in Table \ref{tab:FitTable}. The stellar sample contains only stars with at least
five elemental abundance detections in the range Z$\ge$38. This
guaranteed that the fit had at least three degrees of freedom after
$C_{H}$ and $C_{L}$ were fitted. The total number of stars satisfying
these restrictions combined with the five criteria outlined in
Sect.~\ref{sec:obs_set} is 39 stars.

In order to test if the model is valid, we compare the number of stars
with a $\chi^{2}$ value outside the range, where we expect to find
95\% of the stars (stars with $\chi^{2} \ge \chi^{2}_{5\%} =
2.31$). If that number of stars is much larger than 5\% of the number
of stars considered (in our sample that would correspond to 2 stars
out of 39),   it would indicate that our assumptions are incorrect.  The three models considered are well within
the expected $\chi^{2}$-probability distribution (see
Fig.~\ref{fig:mod6_0.32_5} and Table~\ref{tab:FitTable}).  Only 1
star, \object{CS~22189-009}, is outside the expected range of
$\chi^{2}$ values. Therefore, this star cannot be explained by our
assumption of two robust components. We have also tested   an increment in
the minimum number of elements observed. When this number is increased
to $\ge$9 (reducing the sample to 21 stars), all stars have low
$\chi^{2}$ values.  The good agreement gives credence to the
assumption of two independent robust processes being responsible for
Z$\ge$38 abundances   in most but not all stars.

 The elemental range of the two components is tested in model 3
 (Table~\ref{tab:FitTable}). Here we obtain good fits when using the
 M3-H-component only for Z$>$47 and the M3-L-component only for Z$<$56
 abundances. Without any overlap in the components a good fit is still
 obtained for 38 out of the 39 stars considered (see dotted line in
 the left panel of Fig.~\ref{fig:mod6_0.32_5} and black open squares
 in the right panel). Therefore, using our method it is not possible to
 constrain the elemental range of the components.

We note, that if the criteria of excluding stars with internal mixing
(Sect.~\ref{sec:obs_set}) is removed (model 4,
Table~\ref{tab:FitTable}), the number of stars with $\chi^{2}$ values
larger than $\chi^{2}_{5\%} \ge 2.31$ increases to six stars:
\object{CS~29518-051} (see Fig.~\ref{Fig:x2multi}),
\object{CS~22189-009}, \object{CS~22169-035}, \object{CS~30325-094},
\object{CS~30322-023}, and \object{CS~22783-055}.  Hence, internal
mixing of the giant stars will show a clear effect in the overall
abundance pattern, which is detectable using this method.

   Generally, we obtained good fits to the observations assuming
   two dominant, robust components, but we can neither prove nor
   refute the presence of more processes nor can we at this state
   demonstrate the robustness of the processes. The good $\chi^2$
   indicates that two processes could be sufficient and we discuss
   further in Sect. \ref{sec:scatter} the possible number of primary
   processes and their robustness.

\section{Results and applications of the method}
 \label{sec:results}

%----------------------------------------------------------------------------------------------------
\subsection{The star-to-star scatter}  %3.4
\begin{figure}[!htb]
  \begin{center}
    \includegraphics[width=\linewidth]{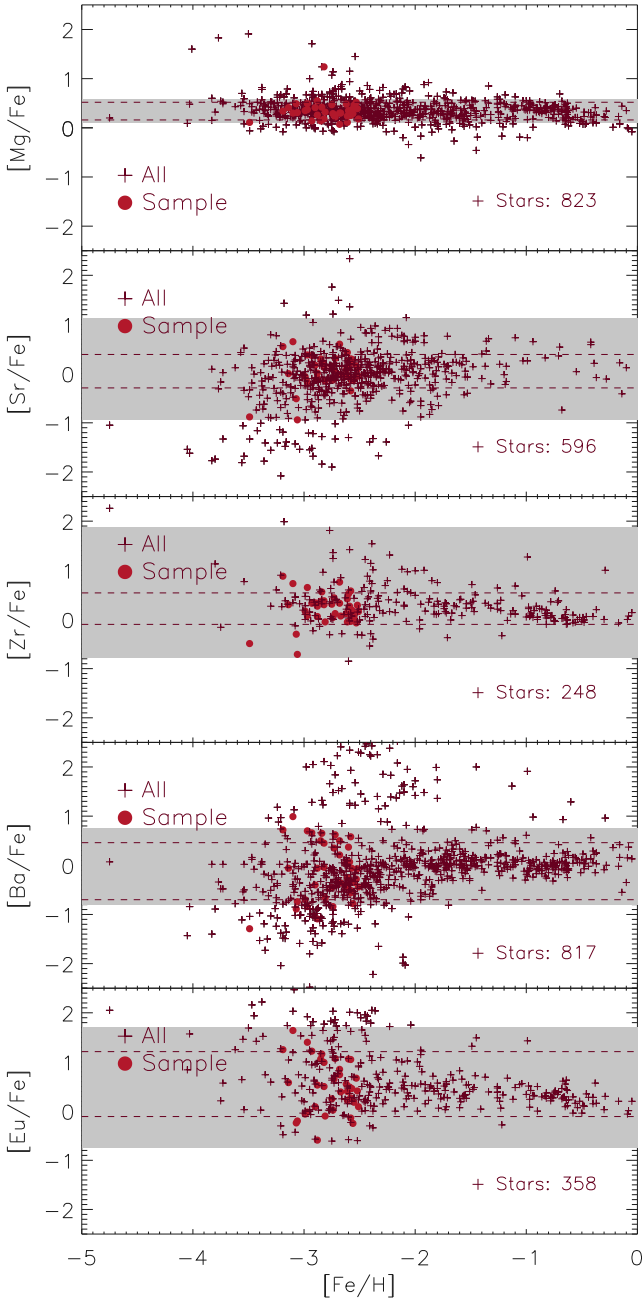}
  \end{center}
  \caption{Top: [Mg/Fe], middle: [Sr/Fe], and [Ba/Fe], bottom: [Eu/Fe]---all plotted versus [Fe/H]. A gray band around the average
    abundance indicates the star-to-star abundance scatter (standard
    deviation) for the full sample, while the dashed lines show the
    standard deviation for our sample. The values are given in Table
    \ref{tab:abunscat}.}
  \label{fig:MgSr_scatt}
\end{figure}

 \label{sec:scatter}
%----------------------------------------------------------------------------------------------------
It is well known from earlier studies \citep[see e.g.,][]{Spite1978,Truran1981,McWilliam1995,Ryan1996,fulbright02,Barklem2005,francois,Roederer2010, hansen12,Yong2013,roederer2014} that a large
star-to-star  abundance scatter as a function of metallicity exists for most
neutron-capture elements.  The large abundance scatter has been
attributed to the elements being produced in more than one
type of nucleosynthesis process or astrophysical environment. For
example, an $\alpha$ element like Mg does not show a large
star-to-star scatter, but only spreads around a mean value of $\sim 0.34 \pm 0.24$\,dex, while many neutron-capture elements show a
scatter of $\pm \sim0.3$ to $1$\,dex or even more (see Fig. \ref{fig:MgSr_scatt} and Table
\ref{tab:abunscat}). This scatter is much too
large to be explained by observational biases or model assumptions
(such as one-dimensional (1D) and LTE).  In this section, we use the derived L  and
H components to split the observationally derived abundances
into their individual contributions and show that even though this scatter
is reduced,   the scatter is still not completely removed in the individual components.

Figure~\ref{fig:MgSr_scatt} shows the abundances of the stellar sample obtained by
applying our selection criteria (Sect.~\ref{sec:obs_set}) compared to the complete
sample as a function of metallicity for a few selected elements. Just by applying
the selection criteria the star-to-star scatter has decreased   (see Table \ref{tab:abunscat}) and the selection
criteria is therefore successful in removing   most of the contamination   originating from multiple processes as well as CEMP stars. 

\begin{table*}[!htb]
\begin{center}
\caption{The mean and standard deviation for our sample and the full un-cut sample.}
\label{tab:abunscat}
\begin{tabular}{lccccc}
\hline
   &  $<$[Mg/Fe]$>$ & $<$[Sr/Fe]$>$ & $<$[Zr/Fe]$>$ & $<$[Ba/Fe]$>$ & $<$[Eu/Fe]$>$ \\
\hline
Sample & 0.34 $\pm$ 0.18 & 0.05 $\pm$ 0.34  & 0.22 $\pm$ 0.32 & -0.12 $\pm$ 0.58  &
0.55  $\pm$ 0.66   \\
Total  & 0.34 $\pm$ 0.24 & 0.10 $\pm$ 1.03  & 0.55 $\pm$ 1.33 & -0.02 $\pm$ 0.78  &
0.47  $\pm$ 1.22 \\
\hline
\end{tabular}
\end{center}
\end{table*}

\begin{figure}[!h]
  \begin{center}
    \includegraphics[width=\linewidth]{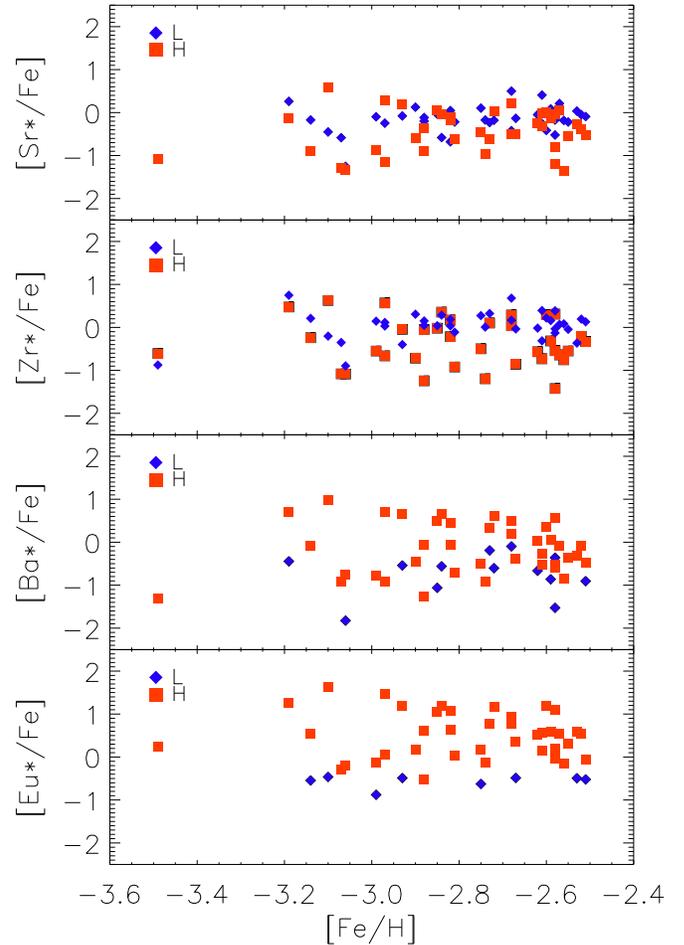}\\
  \end{center}
  \caption{Individual process contribution from the H component and
    L component to [Sr*, Zr*, Ba*, and Eu*/Fe] as a function of [Fe/H] from
    the top to the bottom. The L component is shown as filled blue
    diamonds, while the H component is depicted as filled red
    squares. }
  \label{fig:SrZrBaEu_M2c}
\end{figure}

To fully understand the remaining star-to-star scatter present in the reduced
stellar sample, we rewrite the standard [X/Fe] abundance convention to only include
the contribution from either the L component ([X$_{L}$*/Fe] -- Eq.~\ref{eq:L-componentM2c}) or the
H component ([X$_{H}$*/Fe] -- Eq.~\ref{eq:rM2c}). The first logarithmic term in those
equations corresponds to the total (normal) $\log \epsilon(X)$.

\begin{align}
[X_{L}^{*}/\mathrm{Fe}] =& \log(10^{\log \epsilon(X)} - (Y_{H}(X) \cdot C_{H} \cdot
10^{[\mathrm{Fe/H}]}))  \nonumber \\
 &  -\log \epsilon(X)_{\odot} - [\mathrm{Fe/H}] \label{eq:L-componentM2c}\\
[X_{H}^{*}/\mathrm{Fe}]  =& \log(10^{\log \epsilon(X)} - (Y_{L}(X) \cdot C_{L} \cdot
10^{[\mathrm{Fe/H}]})) \nonumber\\
 & - \log \epsilon(X)_{\odot} - [\mathrm{Fe/H}] \label{eq:rM2c}
\end{align}

\begin{table}[!htb]
\begin{center}
\caption{Mean $\pm$ standard deviation}
\label{tab:mean}
\begin{tabular}{c|cccc}
\hline
      & [Sr$^*$/Fe] & [Zr$^*$/Fe] & [Ba$^*$/Fe] & [Eu$^*$/Fe] \\
\hline
L & $-0.17 \pm 0.32$ & $0.04 \pm 0.33$ & $-1.03 \pm 0.9$ & $-0.46\pm 1.02$\\
H & $-0.44 \pm 0.49$& $-0.36\pm 0.55$ & $-0.15\pm 0.60$ & $ 0.51 \pm 0.54$\\

\hline
\end{tabular}
\end{center}
\end{table}

The L- and H-component contributions to the total abundance are shown in
Fig.~\ref{fig:SrZrBaEu_M2c} as a function of [Fe/H]. Starting with the top panel
showing [Sr$^*$/Fe], we conclude that in most cases the L component contributes the
most to the Sr abundance derived from observations of these stars. Only in a few stars, the H component
contribution is larger.  Table~\ref{tab:mean} shows the mean and standard deviation from the mean for the elements shown in Fig.~\ref{fig:SrZrBaEu_M2c}.  While Zr behaves similar to Sr (L component creating most of the observed abundance), both Ba and Eu are dominated by the H component. A large
scatter is found for Ba and Eu vs. [Fe/H] and it indicates that these two n-capture
H-component elements are not co-produced with Fe \citep[as originally postulated in][]{Spite1978}. 
  The standard deviation increases the more the formation process of the element in the numerator differs from that of the element in the denominator.

  The situation for the lighter elements, Sr and Zr, is less clear
cut because the larger scatter in these elements\footnote{  even the component separated abundances} has the same size as the adopted
uncertainty ($\pm$0.32 dex). Even though definite conclusions are not possible, it should be noted that both \citet{Qian.Wasserburg:2008} and \citet{Li2013} proposed an L-component extending down to iron-peak nuclei, so some co-production of Sr and Zr with Fe is possible. This partial co-production would
reduce the scatter compared to completely uncorrelated nuclei such as Eu.   Hence, this could point toward more than two formation processes contributing to the region ($38<Z<50$) or that the processes are less robust.

\begin{figure}[!htb]
  \begin{center}
    \includegraphics[scale=0.5]{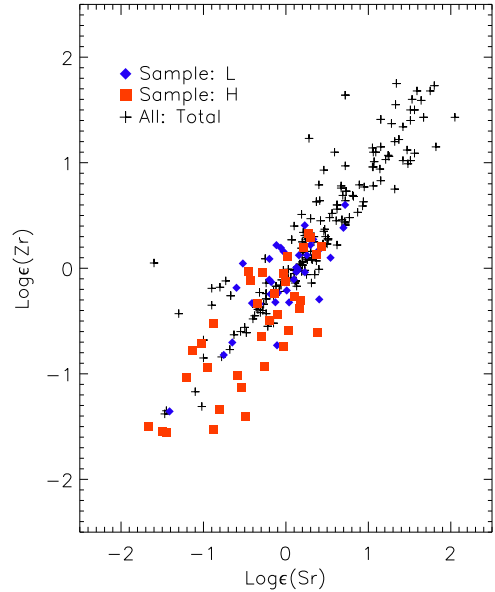}
  \end{center}
  \caption{The calculated contribution from the L component ($Y_L$ blue, diamonds)
    and the main H component ($Y_r$ red, squares) for Sr and Zr.}
  \label{SrZrL-component}
\end{figure}

\begin{figure}[!h]
  \begin{center}
    \includegraphics[scale=0.5]{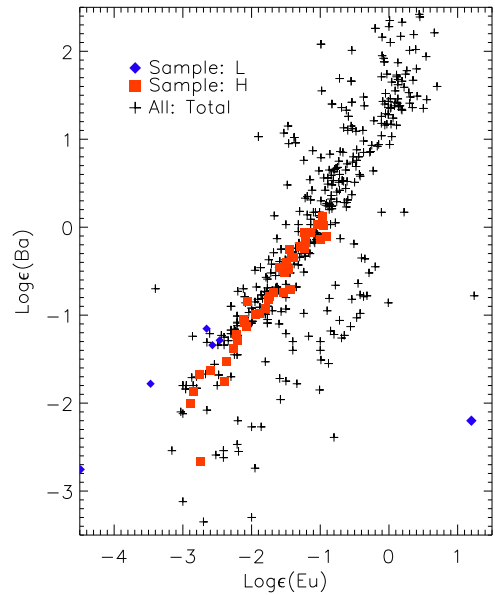}
  \end{center}
  \caption{The L- and H-calculated abundances for Eu and Ba. Details can be found
in the legend.}
  \label{EuBamain}
\end{figure}

The robustness of the L- and H-component contributions can be studied   in
Fig.~\ref{SrZrL-component} and Fig.~\ref{EuBamain}. Barium and Eu show an almost perfect
correlation for the H component with little spread ($\pm0.19$\,dex).   This confirms the robustness of the H component, which is in good agreement with the  universality of the r process (the most likely process behind the H component) shown in many other studies (e.g. \citealt{cowan02,Sneden2003}).
Strontium and Zr, on the other hand,   show a larger star-to-star scatter in both
L  and H components, with standard deviations of 0.27\,dex and 0.38\,dex, respectively.
  Even though the deviations are at the limit of the component uncertainties ($\pm$0.32\,dex), this suggests that
the H component dominates the heavier elements (Z$\ge$56), and that it has a smaller intrinsic
scatter compared to the L component's intrinsic scatter. 
  The relatively speaking larger scatter of the L component may indicate that the component/process is only robust within a $\pm0.32$\,dex uncertainty, or that there may be additional nucleosynthesis components/processes hiding wihtin this uncertainty. The method and current level of abundance accuracy do not allow us to distinguish between these possibilities.

\begin{figure}[!hb]
  \begin{center}
    \includegraphics[width=0.5\textwidth]{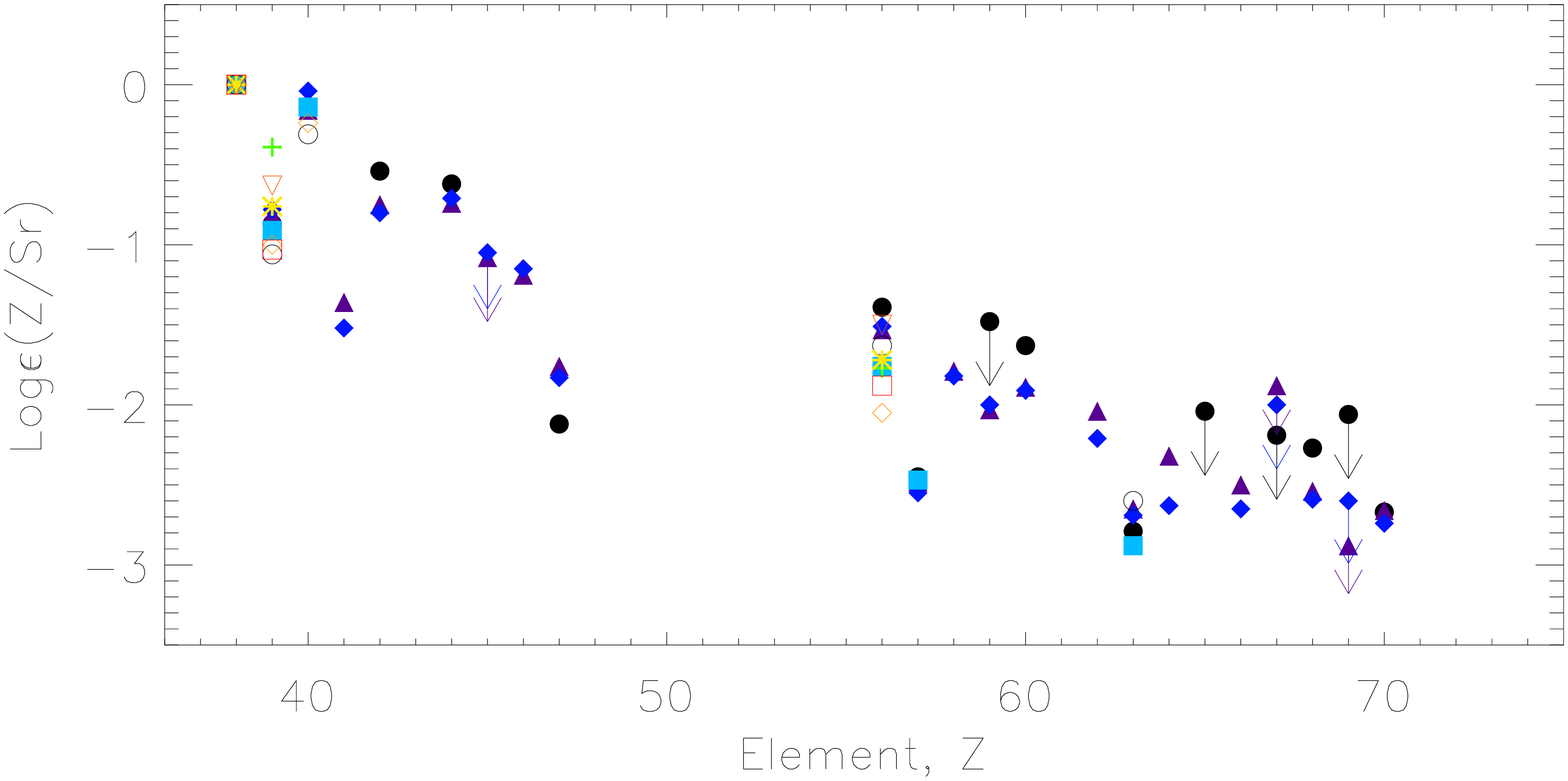}
  \end{center}
  \caption{Abundances normalized to Sr for L-component stars and candidate stars with
promising L-component patterns.}
  \label{fig:L-componentall}
\end{figure}

By extracting stars from our sample with large L-component coefficients, we find
strongly L-component-enriched stars to further test the robustness of the process. 
Figure~\ref{fig:L-componentall} shows the stellar abundances of \object{HE0104-5300},
\object{HE0340-5355}, \object{BS16469-075}, \object{HE1252-075},
\object{HE2219-0713}, \object{BD+4\_2621}, \object{HD4306}, \object{HD88609}, and
\object{HD122563} which have a predominant L-component contribution.  They show a larger
spread in their abundance pattern than first expected (see Fig.~\ref{fig:L-component},
where the well-known L-component stars agree within $\pm 0.2$\,dex   for every element). The larger sample of possible L-dominated stars indicate that these show an abundance spread within
$\pm0.32$\,dex, which is the allowed abundance uncertainty adopted for our pattern
fitting (see Fig.~\ref{fig:L-componentall}).   The increase in the abundance spread could indicate we need to classify the L-component stars such as \object{HD122563} and \object{HD88609} better, to distinguish between such stars and the others shown in Fig. \ref{fig:L-componentall}. Alternatively, the L-dominated stars actually span a wider range of abundances than H-dominated (e.g., the r-II) stars do, and the assumed robustness of the L component may break down. If this is the case, this would allow for a wider range of astrophysical conditions facilitating the process or for several processes to coexist and blend into our L component thereby increasing the scatter to 0.32\,dex.

\begin{figure}[!h]
\begin{center}
\includegraphics[width=0.95\linewidth]{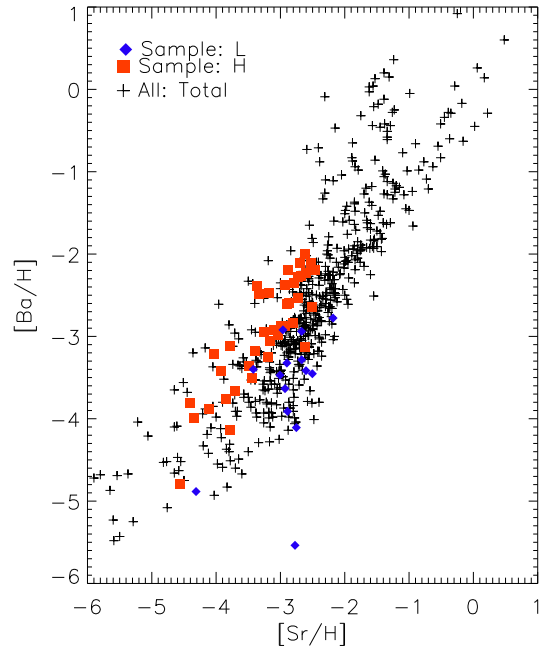}
\caption{Log$\epsilon$ abundances of Sr vs. Ba. The contributions from
  the L component (diamonds) and the H component (main r process) are over
  plotted (squares). \label{SrHBaH} }
\end{center}
\end{figure}

Recently, \citet{ian2013} showed an almost perfect correlation between
Sr and Ba and concluded that all stars must have heavy elements in
their atmospheres.The reason why we do not detect them is
due to weak lines and observational biases. In Fig.~\ref{SrHBaH}, we
show that both Sr and Ba grow (  as in, e.g., \citet{Aoki2005,ian2013}), almost at the
same rate. This could
indicate a coproduction of Sr and Ba in the same site or   that the nucleosynthesis processes
create both elements in almost the same amounts. However, the large star-to-star
scatter could veil differences in the formation process and/or site
between Sr and Ba.   Accurate isotopic abundances for both elements in a large sample are needed to settle this issue. Currently, only Ba has been studied on an isotopic level in small samples (see, e.g., \citealt{roederer2008,gallagher} and references therein). Only when splitting the Sr and Ba abundances into
components we detect the differences in the single formation processes
(for comparison see Figs.~\ref{SrZrL-component} and \ref{EuBamain}). This separation method is currently the best proxy for isotopic abundances. We
clearly see the difference in how Sr is predominantly produced by an
L-component process, a process which does not have to produce any or only
little Ba (see
Figs.~\ref{fig:SrZrBaEu_M2c}--\ref{SrHBaH}). This   separation into L- and H-components partially detaches the
origin of these two elements as they can be produced in different ratios in each component/process. The relation showed in Fig. \ref{SrHBaH}
indicates that despite the fact that all stars have most likely been
enriched in heavy elements, they do not need to be created by the same
process, but could originate from the same object via different
processes   or have been mixed from different sites prior to incorporation. This statement will need to be verified in the future by
improved yield predictions as well as GCE models, which will help us
disentangle the formation sites and   physical quantities.

%----------------------------------------------------------------------------------------------------
\subsection{Predicting abundances}
\label{sec:Eu_predition}
Europium is   an important element because it is almost a pure r-process tracer. 
Thus, we need to know how this element
behaves observationally at the lowest metallicities to constrain and
optimize theory.
\begin{figure}[!h]
  \begin{center}
    \includegraphics[width=0.5\textwidth]{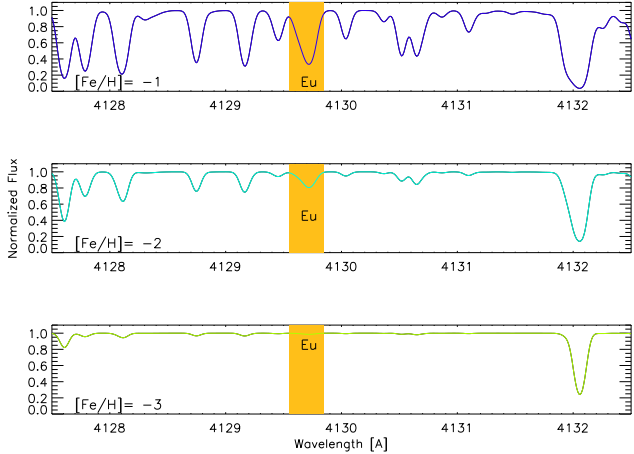}
  \end{center}
  \caption{Three synthetic spectra of giant stars with the same stellar model
    parameters (T=5200K, $\log g $= 2.0), but three different
    metallicities: [Fe/H]= $-1, -2$, and $-3$ in the top, middle, and bottom
    panels, respectively. This shows the effect of metallicity on the
    Eu line strength (still detectable at [Fe/H]$=-2.5$), and the
    difficulties in detecting the very weak Eu lines at [Fe/H]$\leq -3$.}
  \label{fig:Eulines}
\end{figure} 
 Figure~\ref{fig:Eulines} shows that it is very
challenging (or impossible) to measure Eu abundances in extremely and
hyper metal-poor stars (which confirms the bias mentioned in
\citealt{ian2013}). This is a problem because the Eu abundances are
needed to compute abundance patterns and to improve results and
interpretations from, e.g., galactic chemical evolution (GCE) models. These models
rely on the number (statistics) of the observationally derived
abundances. Our method may help improve the statistics for future GCE models of Eu.

\begin{figure}[!h]
\begin{center}
\includegraphics[width=\linewidth]{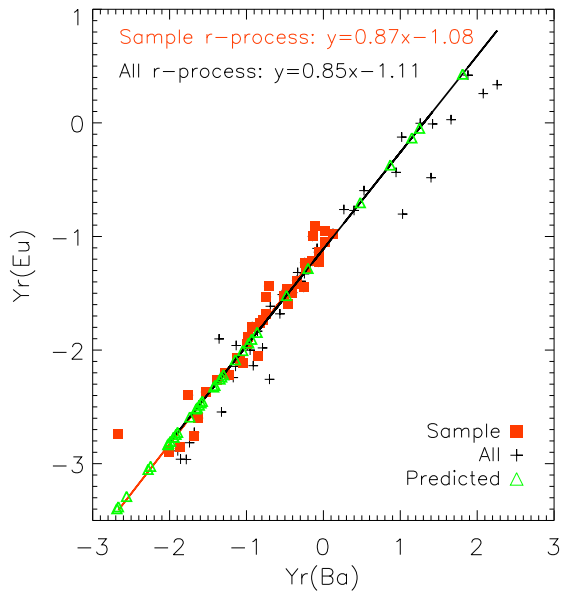}
\caption{H-component fraction of the $\log \epsilon$ abundances
  of Ba and Eu for the selected sample (filled red squared), for all
  stars with calculated components (black plusses), and predicted
  Eu abundances (green triangles). Lines have been fitted to the
  selected sample as well as all stars with calculated components.}
\label{fig:BaEuscaling}
\end{center}
\end{figure}

Since 94\% of Eu is created by the r process/H component \citep[see
  e.g.,][]{Bisterzo2014}, one can use the described method
(Sect.~\ref{sec:components}) to predict Eu abundances, which cannot be
derived from observations owing to the very weak absorption lines found at low-metallicity (or poor quality) spectra. By assuming that Eu is purely created in the H component, Eu can be calculated via a scaling relation between the
H component's Eu and Ba abundances. Barium is an obvious choice
because this element is a good H-component tracer at low metallicity and it
shows fairly strong absorption lines even in extremely metal-poor
stars. Therefore, we know the Ba abundance in a much larger number of
stars than the ones for which we know the Eu abundance (see Fig.~\ref{fig:MgSr_scatt}). 
\begin{figure}[!h]
\begin{center}
\includegraphics[width=\linewidth]{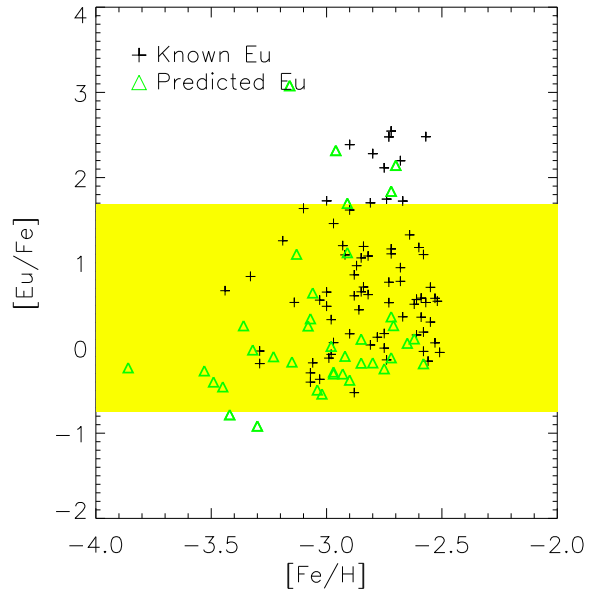}
\caption{[Eu/Fe] as a function of [Fe/H]. (Observationally derived Eu
  shown as pluses.) The predicted, i.e., not observationally detectable
  main H component [Eu/Fe] vs, [Fe/H] (green triangles). The yellow band is the standard deviation as shown in Fig.~\ref{fig:MgSr_scatt}.}
\label{fig:predictedEu}
\end{center}
\end{figure}
The relation
between the H component Eu and Ba (see Fig.~\ref{fig:BaEuscaling}) can
be then used to predict `unobserved' Eu abundances (triangles in the
figure):
\begin{equation}
 \label{eq:predEu}
\log \epsilon({\mathrm Eu})_{r} = 0.87 \cdot \log \epsilon(\mathrm{Ba})_{r} -1.08 .
\end{equation}

By using this relation, the number of Eu abundances known below [Fe/H]
=$-2.5$ would increase by 50\%.  Moreover, the predicted [Eu/Fe]
abundances as a function of metallicity are shown in
Fig.~\ref{fig:predictedEu} and follow the observationally derived
abundances.   This indicates that within the assumption that the method
  produces reasonable Eu abundances, which for most stars are
  trustworthy, but it cannot predict accurate abundances for stars
  such as the outliers mentioned in Sect. \ref{sec:fitting}.

Although, this method can by no means replace real observationally
derived abundances, it can be used to estimate, e.g., Eu abundances either
for GCE calculations or to estimate stellar abundances when applying
for follow-up observing time. 

 %----------------------------------------------------------------------------------------------------
\subsection{Using components to constrain astrophysics}
\label{sec:wind}
%----------------------------------------------------------------------------------------------------

In this section, we use the derived L-component abundances to
constrain the astrophysical conditions of   one of the possible sites
where this component may be produced: neutrino-driven winds in core collapse supernovae. Similar studies have been carried out for the r process (see,
e.g.,~\cite{Mumpower.etal:2012}). 
\begin{figure}[!h]
\begin{center}
\includegraphics[width=\linewidth]{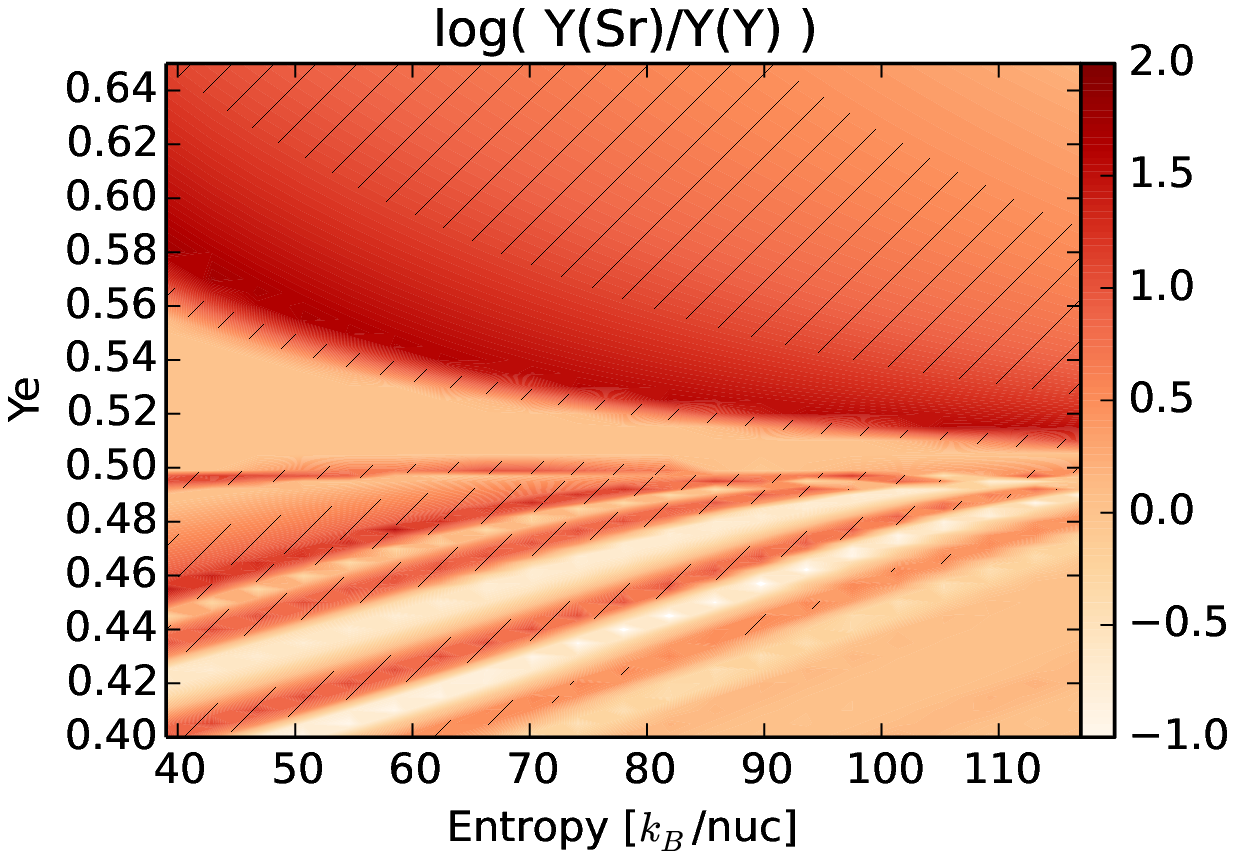}\\
\includegraphics[width=\linewidth]{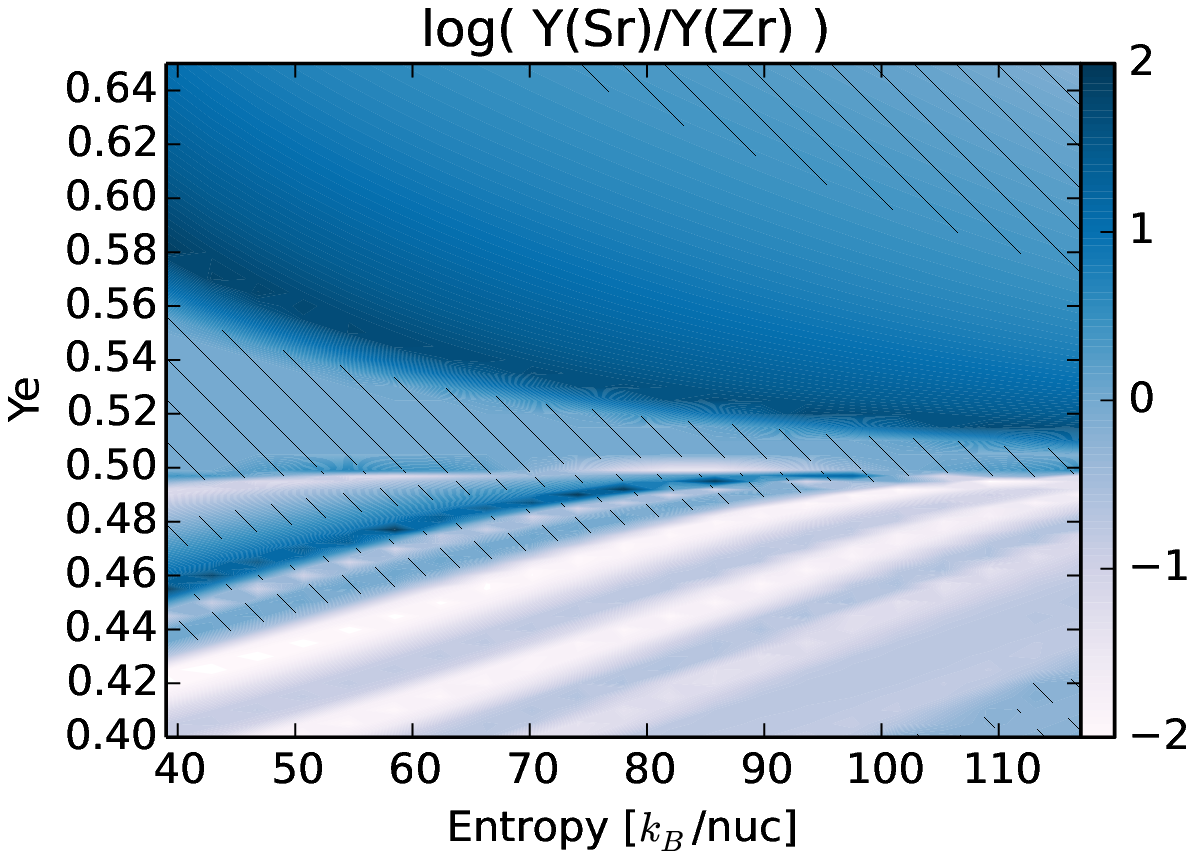}\\
\includegraphics[width=\linewidth]{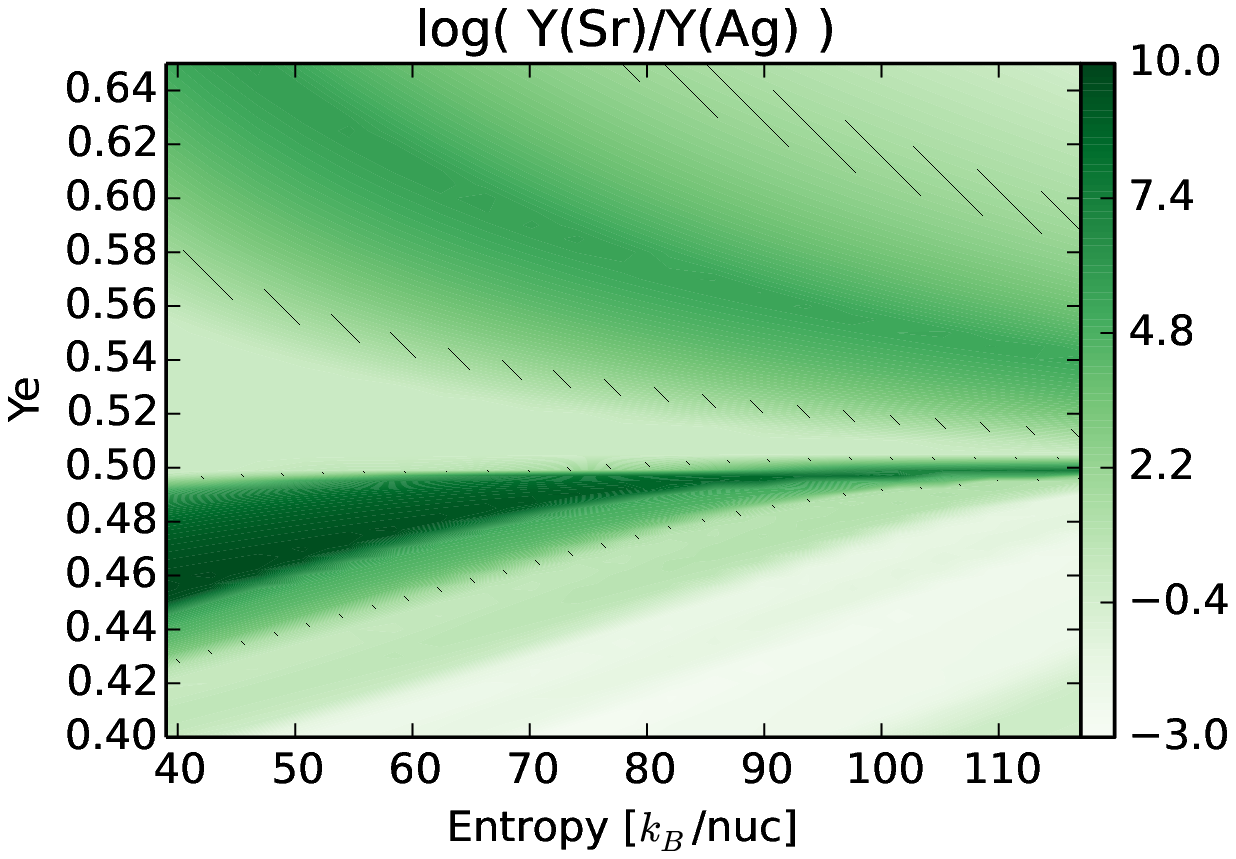}\\
\caption{Ratios of abundances for Sr and Y (upper panel), Sr and
  Zr (middle panel), and Sr and Ag (bottom panel) are shown in
  logarithmic scale for different entropies and $Y_e$. In the hatched
  regions the ratios agree with the ones of the L component $\pm$0.32.}
\label{fig:ratios}
\end{center}
\end{figure}

Neutrino-driven winds occur after a successful core-collapse
supernova explosion, when neutrinos deposit their energy in the outer
layers of the neutron stars, and this layer gets ejected
(see~\cite{arcones.thielemann:2013} for a recent review). Although
neutrino-driven winds were thought to be the site for the
r process \citep{Woosley.etal:1994}, recent hydrodynamic simulations
have shown that the required extreme conditions are not reached. It is
still possible that the winds may have the conditions necessary to
produce the L-component conditions as have been explored in
\cite{Arcones.Montes:2011}.

To explore under which astrophysical conditions in neutrino-driven winds are capable of reproducing the L-component abundances, we have systematically modified a wind trajectory from \cite{arcones.janka.scheck:2007}. This trajectory
corresponds to an explosion of a 15~$M_\odot$ progenitor based
on Newtonian hydrodynamic simulations with a simple neutrino
transport. The simplification in the transport makes it possible to study
the evolution from a few milliseconds up to few seconds post bounce for
various progenitors and explosion energies in one and two dimensions
(for more details see \cite{arcones.janka.scheck:2007,
  Arcones.Janka:2011}). The approximations in the simulations may lead
to small variations of the wind parameters such as expansion timescale,
entropy, and electron fraction compared to what was obtained in the original trajectory. In order to account for the uncertainty in the wind parameters, we systematically varied them within their expected uncertainty. To explore
different entropies ($S\propto T^3/\rho$), the density was reduced and
increased within $\sim \pm 30$\%. The initial electron fraction was
also varied from neutron- to proton-rich conditions. Different expansion timescales were explored by taking slower and faster trajectories obtained in the same simulation at different times.

\begin{figure}[!h]
\begin{center}
\includegraphics[width=\linewidth]{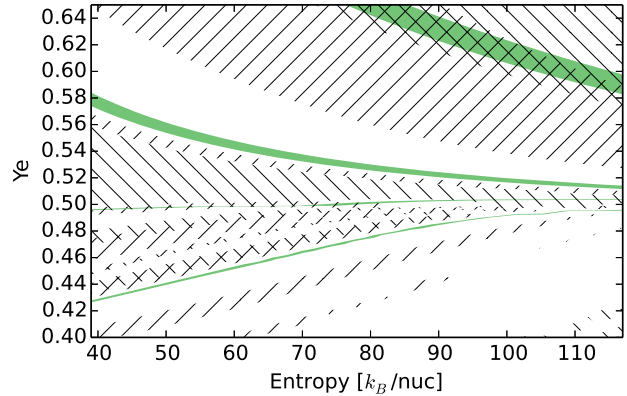}
\caption{L component predicts the ratio of the abundances for
  Sr/Y, Sr/Zr, and Sr/Ag within some error bars. This figure shows the
  wind parameter space and the regions where the ratios Sr/Y ($//$),
  Sr/Zr ($\setminus\setminus$), and Sr/Ag (green) agree with the L-component
  predictions.}
\label{fig:ratios_overlap}
\end{center}
\end{figure}

Figure~\ref{fig:ratios} shows abundance ratios of Sr with respect to
Y, Zr, and Ag using a wind trajectory ejected five seconds after bounce as a
function of entropy and electron fraction.   The calculated L-component ratios shown in Fig. \ref{fig:ratios} are: Sr/Y=6.13, Sr/Zr=1.22, and 
Sr/Ag=48.25. These values with their associated observational and
statistical errors of the method ($\pm0.32$~dex) correspond to the
marked regions in the figures.  Figure~\ref{fig:ratios_overlap} shows
the overlapping regions where all the   simulated ratios (Sr/Y, Sr/Zr, and Sr/Ag)
agree with the   extracted L-component ratios.  For very proton-rich conditions, there is
a band where all ratios overlap. Such conditions ($Y_e\approx 0.62$
and $S\approx 100 k_{\mathrm{B}}/\mathrm{nuc}$) may be achieved a few
seconds after the explosion. Note that the abundances in proton-rich
winds depend on the electron antineutrino luminosity and energy
because the nuclei are produced by the $\nu p$ process
\citep{Pruet.Hoffman.ea:2006,Froehlich06,Wanajo2013}. Slightly
different antineutrino luminosities and energies may result in different entropies and electron fractions. Figure~\ref{fig:ratios_overlap} also shows slightly different proton-rich conditions ($Y_e \sim0.50-0.55$) that could reproduce the derived L-component abundances. However,
this slightly proton-rich ($Y_e \sim 0.50-0.55$) region of the
parameter space is less relevant because the abundances of elements
between Sr and Ag for such conditions are very small.  In neutron-rich
($Y_e<0.5$) conditions the L-component ratios can be reproduced only
in a very narrow band of the parameter space. For small variations of
the wind parameters the abundances change steeply in contrast to the
smooth trend in proton-rich winds (Fig.~\ref{fig:ratios} and
\cite{arcones.bliss:2014}). During the time evolution of the wind, the
parameters are likely to evolve and change. Therefore, given the very narrow range of neutron-rich conditions that can reproduce the L abundances, it is likely that a small change in the astrophysical conditions will result in a non-robust L component. If, on the other hand, the L component is robust (as assumed in this paper), proton-rich conditions are more feasible, since they allow for a wider range of astrophysical conditions and still explain the observed L abundances. 
All previous discussion is based on a single expansion timescale, we have also
studied other trajectories varying this quantity and find that the qualitative behavior and conclusions are the same.

The parameters of the neutrino-driven wind evolve with time as the
neutrino luminosity decreases and the neutron star contracts and
cools. Variations are also expected for different stellar progenitors
because more massive ones will lead to more massive neutron stars which, in
turn, means higher entropy allowing heavier elements to form
\citep{Qian.Woosley:1996, Otsuki.Tagoshi.ea:2000,
  Thompson.Burrows.Meyer:2001}. 
  Therefore, the contribution from
neutrino-driven winds to the observed abundances and to the
L component comes from combinations of wind parameters,
i.e., various points in
Figs.~\ref{fig:ratios}--\ref{fig:ratios_overlap}. In order to explain
the L component, most of the mass needs to be ejected with
parameters from the overlapping regions in
Fig.~\ref{fig:ratios_overlap}, even if any combination of wind
parameters can be realized.  If the contribution from neutrino-driven
winds comes from very different regions of the parameter space, the
abundance pattern would not reproduce the L component and it
would not be robust. Therefore, the robustness (within error bars)
strongly constrains the astrophysical conditions where the L component is produced.

Which is the heaviest element that can be produced in neutrino-driven
winds for typical wind parameters?  
For typical wind
conditions, Ba is not produced or only in a negligible amount.
Therefore, if observations confirm that the L component extends up
to Ba, the neutrino-driven wind (as obtained in current
hydrodynamic simulations) is not likely to be responsible for the L component.

%----------------------------------------------------------------------------------------------------
\section{Summary \& Conclusion}
\label{sec:summ}
%----------------------------------------------------------------------------------------------------
Abundances from metal-poor stars provide clues about the origin of the elements in
the early universe.  We have used the large inhomogeneous sample
of metal-poor stars from \citet{frebel10}, after applying selection criteria to remove contamination from s-process abundances, self mixing, etc., to obtain abundances from single
nucleosynthesis processes. By assuming only two nucleosynthesis processes (the H
and L components following \citealt{Qian.Wasserburg:2001}) contribute to the metal-poor stellar abundances, and that each single event creates robust abundances
(within $\pm0.2$\,dex), we propose that the observationally derived abundances in all metal-poor stars can be explained by a linear superposition of these two contributions.  We have separated the
abundances into single component contributions by different methods (M1, M2, and M3 in Sect. 3.1).
The derived H-component abundances (commonly attributed to the r process) are
remarkably consistent between the different methods (within $\pm0.2$\,dex for every
element). In contrast, even though for elements between Sr and Ag the obtained
L-component abundances are within $\pm0.2$\,dex among different methods, the abundances
vary up to an order of magnitude for elements heavier than Ba. 

For the vast majority of stars in our sample, we have shown that these two robust components are enough to explain the stellar abundance patterns. A similar conclusion was also reached by \citet{Li2013} using a different stellar sample. Most stars have abundances that are well
reproduced by adding up the L and H components, and the abundances of elements
$Z \ge 56$ are generally created by the H component, while the abundance of elements $38 \le Z \le 47$
created by a combination of L and H components. Since the L component does not
have a significant contribution to elements $Z \ge 56$, the exact L abundance to those
elements (obtained by the different methods) does not change or affect the good agreement
between our model and observationally derived abundances. In each method, we found one or a few outlying stars that could not be explained under our assumptions using this method. This could indicate that our method is incomplete or that our assumptions are too strong. However, these outliers (in model 4) seem to have been self-enriched due to internal mixing, which causes the poor $\chi^2$ values.
The general good agreement between our simple
analytical model and the metal-poor stellar abundances indicates that two robust nucleosynthesis processes are responsible for the
abundances of almost all metal-poor stars considered here. 

By deconvolving the stellar abundances into the individual components, we have also
studied the star-to-star abundance scatter. Since the H component is mainly responsible for
the $Z \ge 56$ abundances, most of the observed scatter, as a function of
metallicity is also observed in the H contribution, and it indicates that Fe and the
H component are not co-produced (as originally postulated in \citealt{Spite1978}).
The robustness of the H-component contributions was confirmed by studying the
scatter between different H contributions ($<\pm0.2$\,dex -- see Fig.~\ref{EuBamain}).  For elements
$38 \le Z \le 47$, the situation is not as clear since the observed scatter as a function of
metallicity in the L component is similar to the total attributed
uncertainty in our model ($\pm0.32$\,dex). The scatter between different L-elemental
contributions was also at the limit of what was expected ($\pm0.32$\,dex). This, and the
fact that the abundance pattern of stars with a relative large L-component
contribution is larger than originally expected, suggests that the H component has a
smaller intrinsic scatter compared to the L component's intrinsic scatter. The
relative larger scatter of the L component may either indicate that the process is
robust only within a $\pm0.32$\,dex uncertainty or that there could be additional
nucleosynthesis components or processes hiding within the allowed uncertainty. The
method used in this paper does not allow us to distinguish between these
possibilities.

We consider neutrino-driven winds in core-collapse supernovae as a possible site and use the derived L-component abundances to constrain the astrophysical wind conditions.
In order to explain these abundances, the environment likely needs to be
proton-rich within a significantly constrained parameter space. If the L component is
not robust, or if a single event has evolving conditions, there are even narrower
bands of the parameter space that can reproduce the observations. In addition,
the neutrino-driven winds are not likely to create Ba or heavier elements. If future
abundance observations of L-dominated stars show that Ba is likely to be produced in
an L component, their abundances are not likely to be created in neutrino-driven winds in
core-collapse supernovae.
Both simulations and observations need to improve to further constrain the production of heavy elements and assess the robustness and number of processes working at low metallicity.

 \begin{acknowledgements}
 C. J. Hansen was supported by Sonderforschungsbereich SFB 881 "The
 Milky Way System" (subproject A5) of the German Research Foundation
 (DFG) and the VILLUM Foundation. F. Montes was supported by the Joint
 Institute for Nuclear Astrophysics at MSU under NSF PHY grant
 08-22648. The work of A. Arcones was supported by the
 Helmholtz-University Young Investigator grant
 No. VH-NG-825. We also wish to thank the INT for providing a fruitful forum for discussions (INT-PUB-14-044).
\end{acknowledgements}
\appendix

\section{Uncertainties  from observations and models}
 \label{sec:appendix}
\begin{figure}[!h]
 \begin{center}
 \includegraphics[width=0.6\linewidth]{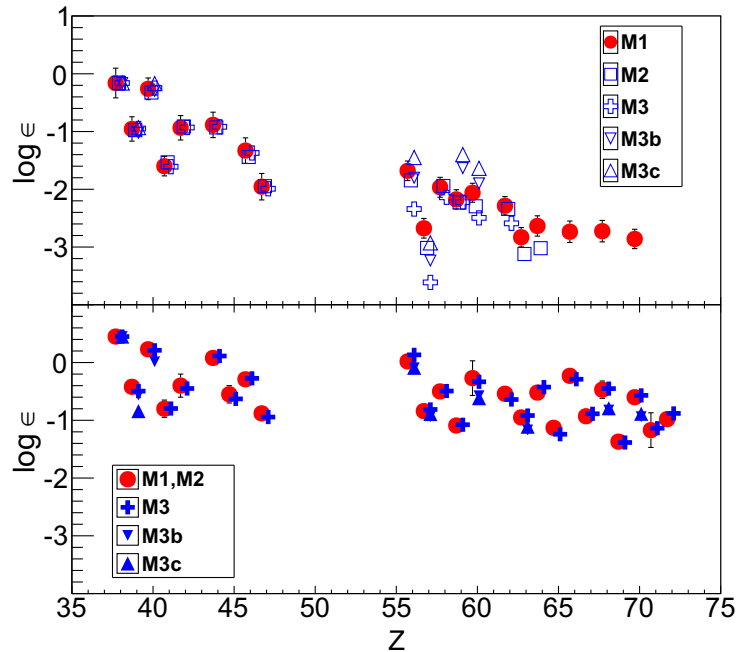}
 \caption{M1, M2, M3 refers to the fitting methods described in
Sect.~\ref{sec:components} and the letters b and c indicate that BD+4\_2621 and
HD6268 were used instead of HD122563 and CS~22892-052.}
 \label{fig:Appuncert}
 \end{center}
 \end{figure}

 To assess how accurate the method is in splitting the abundance contributions from
the L and H components, we carried out several tests to answer two questions: (1) Does
the outcome depend on the intial set of stars? (2) How do the observational
uncertainties (either intrinsic or due to the inhomogeneous stellar sample) affect
the derived abundances? For this purpose, we use a new set of stars that are neither predominantly L-
or H-component enriched, and therefore we used method M3 because it does not assume a dominant
process. \object{BD+4\_2621} and \object{HD6268} were selected because these stars have detailed abundances for a large number of elements. The derivation of the L- and H-component abundances using these stars'
abundances is referred to as method M3b. The impact of the uncertainties on the
observed stellar abundances and the inhomogeneity of the sample was tested by
creating an extreme case based on \object{BD+4\_2621} (method M3c). The new
abundances in this ``modified" star were ``created" by randomly selecting abundances for that specific star found in literature for
each element (generally the literature values vary within the observational
uncertainty). The abundances of other stars were not modified.
Figure~\ref{fig:Appuncert} shows the abundances obtained by using \object{BD+4\_2621} and \object{HD6268} in method M3b, and using method M3c for the ``modified" \object{BD+4\_2621} and \object{HD6268}. Similar
results are obtained using different combinations of stars. All H- and L-component
abundances in the range $38 \le Z \le 47$ are very similar and show a good
agreement within $\pm$0.2\,dex.  By comparing the results of methods M3b and M3c,
it can be seen that the uncertainty in the abundance determination due to the
observational uncertainty is on average within  $\pm$0.25\,dex   for all elements. For L-component nuclei with $Z\ge 56$,
the differences among the methods (M3, M3b, and M3c) are larger and this reflects
the intrinsic limitations of the method (or physical features of the process, since the L component may not create these nuclei very efficiently). The more similar the abundances of the stars
are, the larger the uncertainty in the decomposed abundances get. 
The difference between the methods can be taken as a conservative uncertainty in the
component abundances, because \object{BD+4\_2621} are \object{HD6268} are not dominated by an individual component (L or H). The uncertainty obtained based on M3, M3b , and M3c is
similar to the one based on M1, M2, and M3. Therefore, the use of different L-component
abundances (M1 and M2 in Table 1) also covers the error due to the choice of the
initial stars. 
 
Since the uncertainty for both components (excluding L component $Z\ge 56$) is within
$\sim 0.2$\,dex and the observationally derived abundances have on average 0.25\,dex
uncertainty, these contributions were added in quadrature, yielding a value of $\pm 0.32$\,dex, and it is used in the abundance deconvolution described in the Sect.~\ref{sec:fitting}.  

\bibliographystyle{apj}
\bibliography{bibliography}

\end{document}